\documentclass{aa}
\usepackage{times}
\usepackage{paralist}
\usepackage{float}
\usepackage{array}
\usepackage{tabularx}
\usepackage{changepage}
\usepackage{color}
\usepackage{graphicx}
\usepackage{natbib}
\usepackage{mathabx}
\usepackage{multirow}
\usepackage{txfonts}
\usepackage{bm}

\def\degr{\hbox{$^\circ$}}
\def\arcmin{\hbox{$^\prime$}}

\begin{document}

   \title{The NOD3 software package: A graphical user interface-supported reduction package for single-dish radio continuum and polarisation observations}
   \titlerunning{NOD3:\ A GUI supported reduction package}


   \author{Peter M\"uller\inst{1}, Marita Krause\inst{1}, Rainer Beck\inst{1}, and Philip Schmidt\inst{1}}
   \authorrunning{Peter M\"uller et al.}

   \institute{Max-Planck-Institut f\"ur Radioastronomie,
              Auf dem H\"ugel 69, 53121 Bonn, Germany\\
              \email{peter@mpifr-bonn.mpg.de}
             }

   \date{Received 2017 May 22; accepted 2017 Jul 13}


  \abstract
   {The venerable NOD2 data reduction software package for single-dish radio continuum observations,
   which was developed for use at the 100-m Effelsberg radio telescope, has been successfully applied over many decades.
   Modern computing facilities, however, call for a new design.}
   {We aim to develop an interactive software tool with a graphical user interface (GUI) for the reduction of single-dish radio continuum maps. We make a special 
   effort to reduce the distortions along the scanning direction (scanning effects) by combining maps scanned in orthogonal directions or
   dual- or multiple-horn observations that need to be processed in a restoration procedure. The package should also process polarisation data and 
   offer the possibility to include special tasks written by the individual user.
   }
   {Based on the ideas of the NOD2 package we developed NOD3, which includes all necessary tasks from the raw maps to the final maps 
   in total intensity and linear polarisation. 
   Furthermore, plot routines and several methods for map analysis are available.
   The NOD3 package is written in Python, which allows the extension of the package via additional tasks.
   The required data format for the input maps is FITS.
   }
   {The NOD3 package is a sophisticated tool to process and analyse maps from single-dish observations that are affected by scanning effects
   from clouds, receiver instabilities, or radio-frequency interference (RFI).
   The 'basket-weaving' tool combines orthogonally scanned maps into a final map that is almost free of scanning effects.
   The new restoration tool for dual-beam observations reduces the noise by a factor of about two compared to the NOD2 version.
   Combining single-dish with interferometer data in the map plane ensures the full recovery of the total flux density.
   }
   {This software package is available under the open source license GPL for free use at other single-dish radio telescopes of the
   astronomical community. The NOD3 package is designed to be extendable to multi-channel data represented by data
   cubes in Stokes I, Q, and U.
   }


   \maketitle
%

\section{Introduction}

Radio continuum and polarisation data from the 100-m Effelsberg radio telescope have usually been reduced
and analysed with the NOD2 reduction system \cite{haslam74} and the library of routines described in the
Hitch-Hiker's Guide to NOD2 \cite{andernach85}. The programs were written in Fortran and the maps needed
to be in a special NOD2 data format. Several procedures were developed within NOD2 to reduce the
so-called scanning effects:

Scanning effects in single maps can be suppressed with 'presse' \cite{sofue+79}.
Maps observed with multi-horn receivers can be combined with a restoration software \cite{emerson+79}.
Maps of observations scanned along different orientations within the sky can be combined with
'basket weaving' (method unpublished) for orthogonally scanned maps or 'plait' for double- or multi-horn
observations \cite{emerson+88}.

Present-day computing facilities offer interactive usage and graphical displays, fast data processing,
and huge data storage. We designed an interactive software package for single-dish radio continuum observations, based on the FITS format, 
named NOD3, with a graphical user interface (GUI) that includes all tasks to produce final maps in
total intensity and linear polarisation. Several of the data reduction methods originally introduced by NOD2
were significantly improved or revised.

While the widely used Common Astronomy Software Applications (CASA) package \cite{mcmullin+07} is
mainly designed for the reduction and analysis of interferometer data,
the new NOD3 package is -- similar to its precursor NOD2 -- explicitly developed for the different requirements of single-dish 
radio continuum observations.

\section{Data format and general features}

In order to make the new software package, NOD3, as flexible and compatible as possible with observations from
other telescopes we chose the  Flexible Image Transport System (FITS) format as the data format for the input maps.
We labeled the header parameters of the FITS format identical to those chosen in CASA \cite{mcmullin+07}.
Additional header information can easily be added.

The program package is written in Python, which comfortably allows the user the inclusion of special tasks written by the
individual user. In contrast to NOD2, the astronomical data in NOD3 are represented as a two-dimensional array
or as a data cube for multi-channel observations, one each in Stokes I, Q, and U with its own header parameters.

The input parameters for all tasks can be entered either with the cursor or directly by typing the value.
These parameters are shown and also stored within the tasks for the next application. They are also automatically compiled
in a logfile that can easily be used to generate scripts for future utilisation and also for sequences (scripts)
of different tasks. All tasks that have been applied to a map are listed in the NOD3 history file following the
FITS header.

\section{Typical data reduction steps for single-dish observations}

The data stream (time series) from a telescope needs to be transformed into a two-dimensional map. We recommend the following steps:
\begin{itemize}

\item Filtering of radio-frequency interference (RFI) in the raw continuum data stream from the output channels of the polarimeter
(Stokes I, Q, U for each horn).

\item Calibration of the Stokes parameters (I, U, Q) by applying the $Mueller\, Matrix.$ 

\item Interpolation of the raw data with a $sinc$ weighting function into square pixels and combination into maps in FITS format.

\item For maps scanned in azimuth direction, generate two additional maps for the sidereal time and the parallactic angle for each pixel in the map, to allow the transformation from the horizontal 
coordinate system to the equatorial coordinate system, if necessary.

\end{itemize}

For each radio telescope a specific system for generating the raw data has been developed.
Those of the 100-m Effelsberg telescope are preprocessed with the 'Toolbox' package (see Appendix).
The format of the raw data from multi-horn Effelsberg observations is set in a way that all maps
(Stokes I, U, Q) of a single coverage of a target are stored in one file, together with two additional maps
for the sidereal time and parallactic angle for each pixel in the map.

The NOD3 package includes all necessary steps for the reduction of the raw FITS maps to the final maps.
First, it is necessary to correct the base levels and suppress the scanning effects. 
These scanning effects can be further reduced by the following reduction steps:

\begin{itemize}

\item Combination of maps (for each channel) from multi-horn observations

\item Averaging of the maps of total intensity channels to a map in Stokes I

\item Combination of all single coverages to maps in Stokes I, Q, and U

\item Computation of maps of linearly polarised intensity PI (bias-corrected) and polarisation angle PA

\item Calibration of absolute flux density with help of a map of a calibration source

\item Checking and correction of absolute polarisation angle with the help of a map of a linearly polarised calibration source

\end{itemize}

Further, NOD3 offers plot routines and several methods for the map analysis such as an integration in sectors of
elliptical rings in maps of total intensity and polarisation (e.g. in the projected plane of an inclined galaxy) and box integration along strips including scale height determinations (e.g. in edge-on galaxies).

\section{Improved methods}

Although we kept the original ideas of the reduction routines from the NOD2 library, we refined the methods
if possible (e.g. $Flatten$) and also introduced some new procedures (e.g. $Restore$). As a result, we are able
to reduce the scanning effects, which are distortions in the direction along which the map area was scanned on the
sky during the observations. Furthermore, we can significantly reduce the root mean square (rms) noise in the maps.

\subsection{Removal of scanning effects, base level correction, and noise reduction}

\subsubsection{Base level correction: {\it Adjust}}

The measured system temperature of the receiver not only contains astronomical signals but also contains atmospheric and
ground radiation. The latter depends on the elevation of the telescope and is usually much larger than the astronomical
signal. The base level can be corrected for by a polynomial fit to each scan of a map. For small
scanning lengths ($< 1^\circ$) it is sufficient to correct the base level linearly, otherwise the fit can have a
higher order.

In principle we do not know the absolute base level of a map. Hence, we have to adjust the base level relative to zero at the beginning
and at the end of each scan. To achieve this, we select a range of points on each side of each scan over which we determine the
base level by a median filter. Although we get different median values for each scan, we get smooth transitions to
adjacent scans owing to the application of the median filter. The advantage of the median filter is that it excludes extrema
caused by point sources or RFI. Base levels can be affected by compact astronomical sources at the beginning or end of a scan, which
is avoided with our filtering method.
If scanning effects still remain, other methods are required; see e.g. Sect.~\ref{flatten}.

\subsubsection{Suppressing of scanning effects: {\it Flatten}}
\label{flatten}

Because of several kinds of scanning effects (i.e. weather, atmospheric or ground radiation, and receiver instabilities)
an individual single map usually still includes scanning effects after adjusting the scans to a zero level.
It is possible to reduce these distortions with the method of ``unsharp masking`` \cite{sofue+79}. In contrast to that method, we do not smooth the data to values larger than the beam width of the telescope along the scanning direction
and only slightly perpendicular to the scanning direction\footnote{According to the sampling theorem, the separation between scans must be smaller
than half the beam width.} by 1.5 times the scan separation;
this ensures that astronomical structures are not affected. 
The difference between the original and smoothed map is taken to fit a polynomial baseline iteratively.
An acceptable result is reached after a few iterations, for which fits of the order 3 or higher can be applied if necessary.

\subsubsection{Single-horn observations: \emph{BasketWeaving}}\label{sect:basket}

This task can be used to remove scanning effects in maps that
are scanned at least once along two orthogonal directions (in astronomical coordinates).
Clouds moving in the atmosphere cause a time-dependent base level, which shows up as scanning effects.
Base level distortions due to atmospheric or ground radiation cannot be removed and can only be suppressed by averaging. If more than one coverage is available in each
scanning direction, all maps with the same scanning direction are averaged first.

\begin{figure}[h]
 \includegraphics[scale=0.6,keepaspectratio=true,clip=true,trim=30pt 0pt 0pt 0pt]{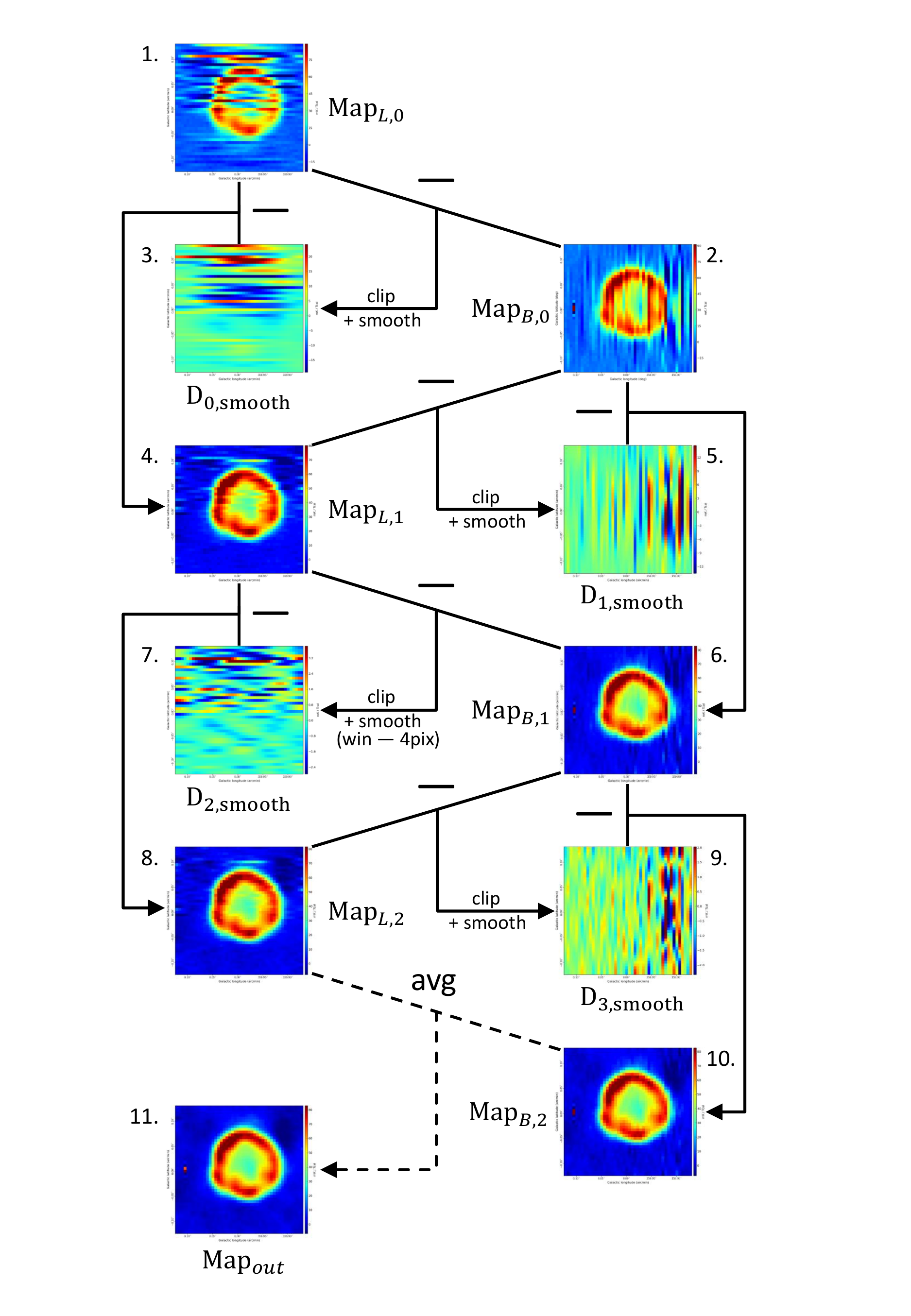}
 \caption{Showing the work flow of the basket-weaving procedure, based on observations of the Tycho supernova remnant at 4.85 GHz with the Effelsberg 100-m telescope. 
 The sketch shows 2 iterations of the method.}
 \label{bweav}
\end{figure}

The \emph{BasketWeaving} algorithm is illustrated in Fig.~\ref{bweav}. The goal is to remove the scanning effects by iteratively forming difference maps between
the two orthogonal maps, starting with the input maps $Map_{L,0}(l,b)$ and $Map_{B,0}(l,b)$ (panels 1 and 2 of Fig.~\ref{bweav}, respectively),
\begin{equation}
 D_{0}(l,b)=Map_{L,0}(l,b)-Map_{B,0}(l,b).
\end{equation}
This should remove the astronomical signal completely, except in case of pointing errors. Next, the global rms value of the difference map
is calculated and all map values greater in magnitude than $\pm\sigma_{\textrm{clip}}\times rms$, where $\sigma_{\textrm{clip}}$ is a user-given input, are set to this value.
To ensure that no astronomical signal is left in the difference map, this cycle of rms computation and clipping is iterated three times. Each row of the clipped
difference map is then smoothed by a (one-dimensional) Hanning window of width corresponding to half the column or row length, and the smoothed map (Fig.~\ref{bweav}, panel 3)
is subtracted from the original (averaged) map in the according scanning direction
\begin{equation}
 Map_{L,1}(l,b)=Map_{L,0}(l,b)-D_{0,smooth}(l,b)
.\end{equation}
The resulting map (panel 4), in turn, is now subtracted from the original (averaged) map in the other scanning direction
\begin{equation}
 D_{1}(l,b)=Map_{B,0}(l,b)-Map_{L,1}(l,b),
\end{equation}
and the above steps (clipping, smoothing, and subtraction of the smoothed map) are repeated, but this time the Hanning smoothing is performed for each column
(panels 5 and 6), i.e.
\begin{equation}
 Map_{B,1}(l,b)=Map_{B,0}(l,b)-D_{1,smooth}(l,b)
.\end{equation}
The entire process is iterated for successively narrower smoothing windows, hence higher spatial frequencies, 
until a minimum window size is reached. We choose a step size of four pixels and a minimum window size of five pixels, but these values are not critical. 

In a final step (panel 11), the program forms the average of the last maps obtained in the two orthogonal scanning directions, which should be ``cleaned'' from
scanning effects on various length scales,
\begin{equation}
 Map_{out}(l,b)=Map_{L,final}(l,b)-Map_{B,final}(l,b).
\end{equation}

This \emph{BasketWeaving} method is effective even for a pair of orthogonally scanned maps.

\subsubsection{Restoration of dual- or multiple-horn observations: {\it Restore}}

Dual- or multiple-horn observations can reduce scanning effects significantly. If the horns are
mounted parallel to the azimuth direction and the telescope covers the field of the sky with
azimuth scans, the assumption is that the horns detect the same clouds but at different positions on the sky.
Emerson et al. (1979) published a method to restore multiple-horn observations obtained with a single-dish radio telescope.
This method has the disadvantage that the noise increases with the number of pixels along a scan,
which was noted already by Emerson et al. (1979). The method uses the
difference of two maps covered by different horns with the underlying idea that the weather effects disappear while
the astronomical signal remains in difference, shifted by the horn separation.

\begin{figure}[h]
 \includegraphics[scale=0.185,keepaspectratio=true,clip=true,trim=30pt 0pt 0pt 0pt]{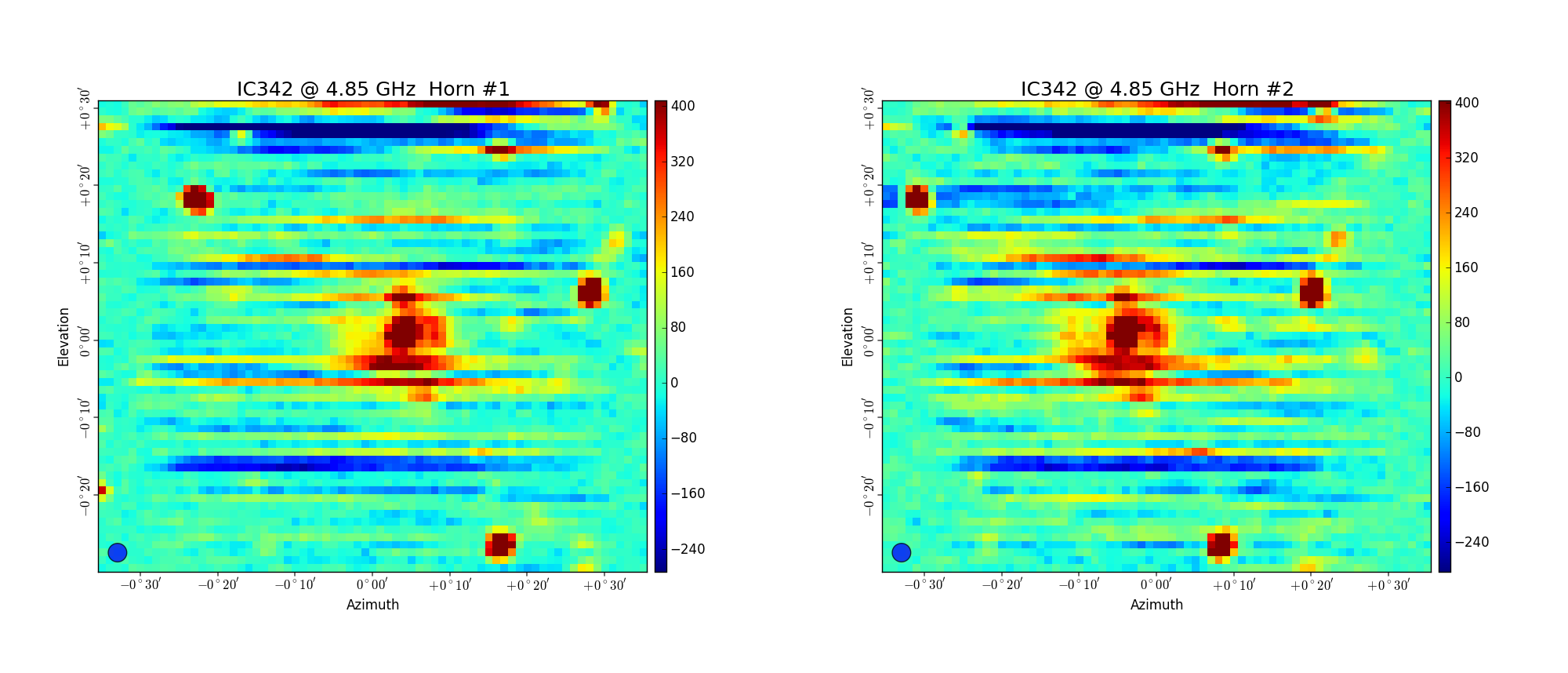}
 \caption{One coverage of the spiral galaxy IC~342 observed at 4.85~GHz with the Effelsberg 100-m telescope \cite{beck15}, simultaneously covered by the two different horns
 in the secondary focus.
  The maps from the two horns show an offset in source position, while the weather effects are at the same position.}
 \label{restore12}
\end{figure}

\begin{figure}[h]
 \includegraphics[scale=0.185,keepaspectratio=true,clip=true,trim=30pt 0pt 0pt 0pt]{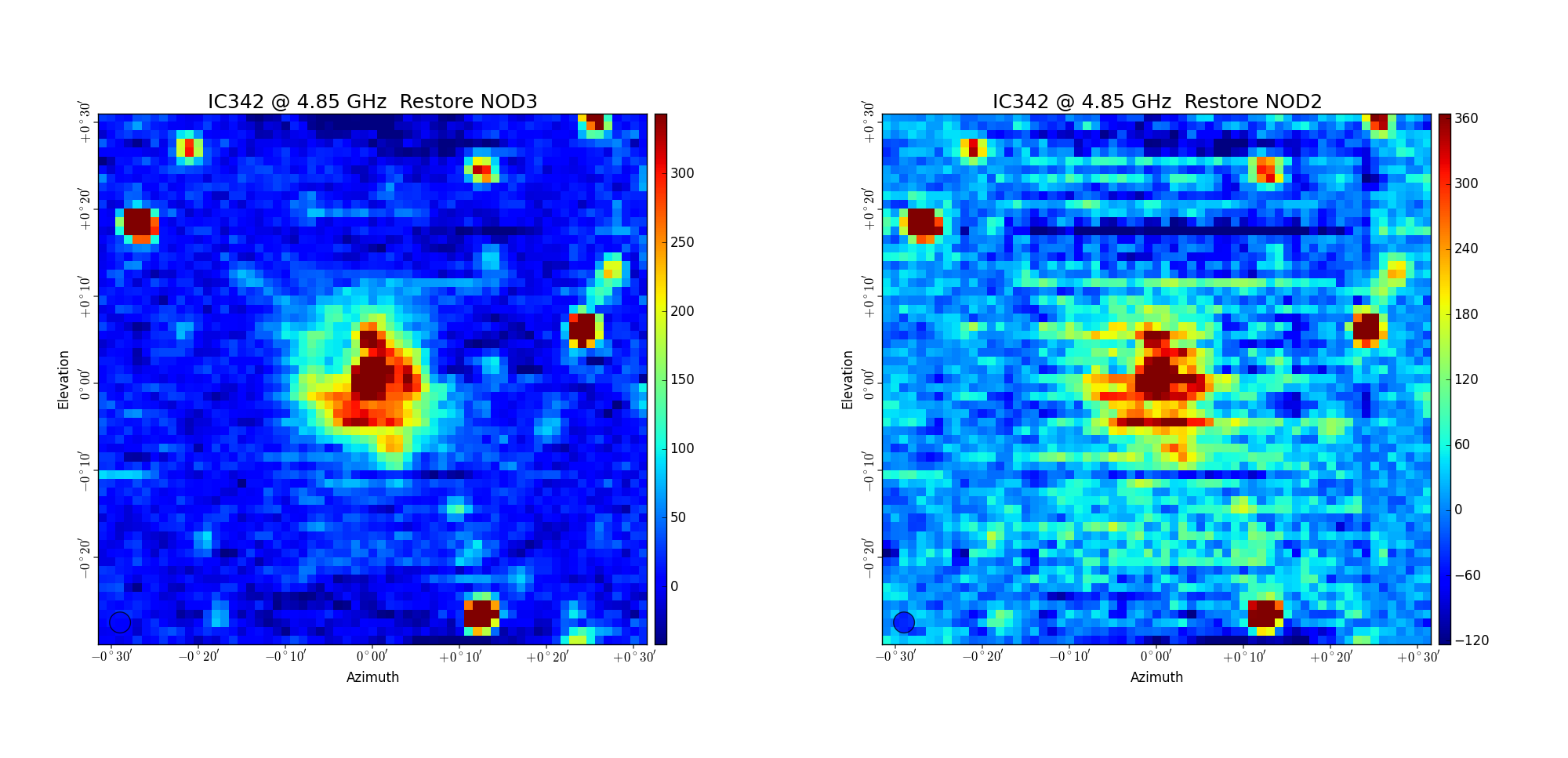}
 \caption{Comparison of the results of the restoration process of the two maps shown in Fig.~\ref{restore12} after applying the NOD3 method (left) and NOD2 method (right).}
 \label{restoreNOD}
\end{figure}

\begin{figure}[h]
 \includegraphics[scale=0.185,keepaspectratio=true,clip=true,trim=30pt 0pt 0pt 0pt]{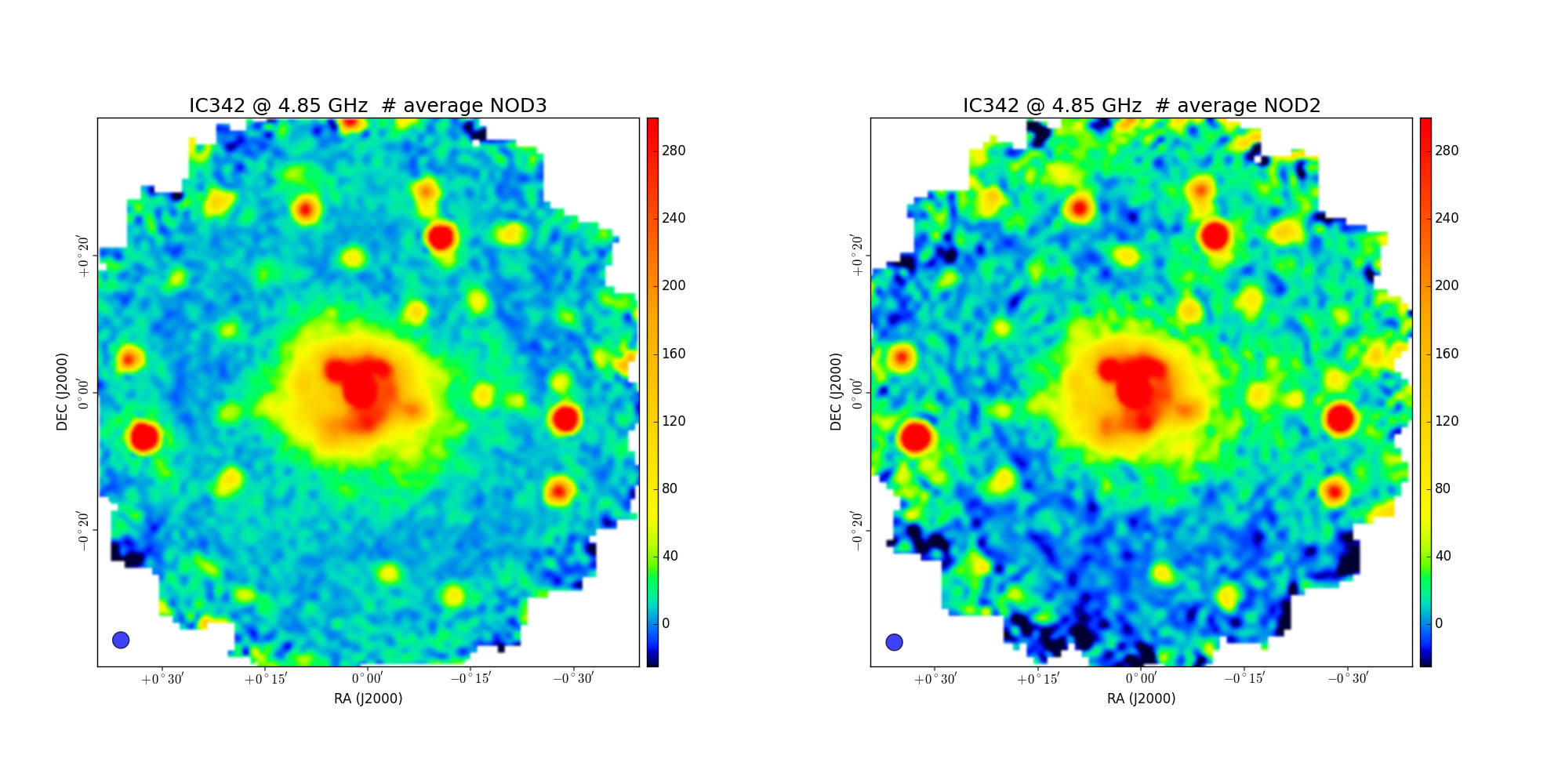}
 \caption{Comparison of the results of the restoration process of all 20 coverages observed by Beck (2015) after the NOD3 method (left) and NOD2 method (right).}
 \label{restoreIC342}
\end{figure}

The new method takes a different approach by shifting the related scans of two horns by their azimuth
distance. In this case, when computing the difference map, the astronomical signal disappears but the scanning
effects due to the weather are now seen in difference. So we can manipulate the difference map without running the
risk of influencing the astronomical signal.

If $x$ is the astronomical position (measured in azimuth) at the sky and $h$ is the distance in azimuth of
two horns $H1$ and $H2$, then

\begin{equation}
 D(x) = [H1(x+h/2) - H2(x-h/2)] / h \, ,
\end{equation}
corresponds to a derivation of the signals of the two horns (Fig.~\ref{restore12}), assuming equal weather effects in both
horns. Now, an integration of the function $D(x)$ solves the problem and we can easily restore the scanning
(weather) effects $W(a)$ at azimuth position $a$,

\begin{equation}
 W(a) = \int_{0}^{a} D(x) \cdot dx + c \, ,
\end{equation}
and subtract it from the corresponding scans from each of the horns. The integration constant $c$ yields an offset, which has to be corrected separately with the $Adjust$ method described in section 4.1.1.

In our case $h$ describes the distance of two horns.\ The value $h$ is always greater than zero. For a mathematically correct treatment $h \rightarrow 0$ is required.
This acts like a $low-pass$ filter and structures on scales smaller than the distance of the horns are not detectable. Small structures, however, are not relevant
in our case, as we assumed that weather signals on small scales are equal in both horns, hence its angular
extent is larger than that corresponding to the horn separation.

The low-pass filter has a positive effect as it avoids significant noise in the restored weather effects
$W(a)$ at azimuth position $a$. After subtracting $W$ from the corresponding scans of the two horns, averaging of
the scans leads to a noise improvement by a factor of $\sqrt{2}$ compared to that in single-horn
observations, regardless of the length of the scan.

The noise levels in Fig.~\ref{restoreNOD} are 14.7 units (left) and 29.3 units (right). So the signal-to-noise ratio of the new restoring method is two
times better than for the old NOD2 method. The old NOD2 method increases the noise with the number of pixels per azimuth row. The new NOD3 method does not
increase the noise but corrects the
weather effects in each map separately, which yields an improvement in total of about a factor of {2}.

Background sources at the edges of the scanned area do not influence the restoration process with the new method because only the overlapping
areas are considered and hence do not have to be eliminated beforehand. This is different from the previous method of restoring double-horn observations
by Emerson et al. (1979).
Furthermore, the maps created with the new method do not suffer from a maximum visible structure
of about three~times the horn distance as was the case for the previous method by Emerson et al. (1979).

If a large number of coverages are combined, the final map created with the new method (Fig.~\ref{restoreIC342} left)
shows clear improvements compared to the map created with the previous method (Fig.~\ref{restoreIC342} right). That is, the improvement
in the noise level is a factor of 2, the background level is smoother, and fewer artefacts are visible at the edges. As a
result, the diffuse emission of the galaxy can be traced to larger radii. A level of three~times the rms noise is reached at
43\arcmin\ radius compared to 35\arcmin\ radius with the previous method \cite{beck15}.

For observations with more than two horns, it is possible to combine all
pairs of horns that are aligned in the azimuth scanning direction. However, the distance of the combined horns
should not be too large, so that weather effects are similar in all horns and the low-pass filter effect is not
too large.

\subsection{Polarisation: {\it PolInt}}

Polarised radio emission comes as Stokes $Q$ and $U$ maps. Both contain receiver
noise $N_{Q,U,P}$ that is not part of the astronomical signal. The noise shifts the observed polarised intensity
$\hat{P} = \sqrt{(Q_T+N_Q)^2 + (U_T+N_U)^2\ }$ up with respect to the 'true' polarised intensity $P_T$, which  is called the polarisation bias.

The difference $\hat{P} - P_T$ has the probability density known as the $Rice \ distribution$, which is always
positive definite. The bias effect becomes more significant for smaller signal to noise of $Q_T$ and $U_T$.
Rewriting the above formula to
\begin{equation}
 \begin{array}{ll}
  P_T + N_P = (U_T+N_U)\sin(\theta_T) + (Q_T+N_Q)\cos(\theta_T) \, ,\\
  \theta_T = \tan^{-1}(U/Q) \, ,
\end{array}
\end{equation}
allows us to get rid of the bias. The method is described comprehensively by M\"uller et al. (2017).
It requires the knowledge of the 'true' angle $\theta_T$ from the 'true'
$Q_T$ and $U_T$. A good approximation is given by the 'modified median filter' of the adjacent angles in the map as described in M\"uller et al. (2017).

In most cases we have just $Q$ and $U$ and need to build the modified median value $\theta_m$ of the adjacent angles,

\begin{equation}
 \begin{array}{rcl}
  P=P_T + N_P &=& (U_T+N_U)\sin(\theta_m) + (Q_T+N_Q)\cos(\theta_m) \, ,\\
  PA &=& \frac{1}{2} \theta_m \, ,\\
 \end{array}
\end{equation}
where $PA$ is close to the 'true' polarisation angle. Numerical simulations show that $P$ is almost
bias free and the probability density of the noise $N_P$ is Gaussian with the same standard deviation as those of
$N_Q$ and $N_U$. The noise distribution of $N_Q$ and $N_U$ does not have to be Gaussian but needs to be
centred around zero.

\subsection{Combination of single-dish with interferometer observations: {\it Immerge}}

It is a well-known problem that large holes in the u,v plane of interferometric data
causes 'missing flux' because telescope arrays do not cover all short baselines. This missing flux
can be added from a large single-dish such as the Effelsberg 100-m radio telescope. Ideally, the receiver frequencies
should cover the same range at both observations. Instead of combining both maps in the Fourier plane as in
AIPS \cite{greisen03}) or CASA \cite{mcmullin+07}, we propose a straightforward combination method in the image plane. The resulting map is
similar to the result of ``feathering'' in CASA.

\begin{figure}[h]
 \includegraphics[scale=0.15,keepaspectratio=true,clip=true,trim=30pt 0pt 0pt 0pt]{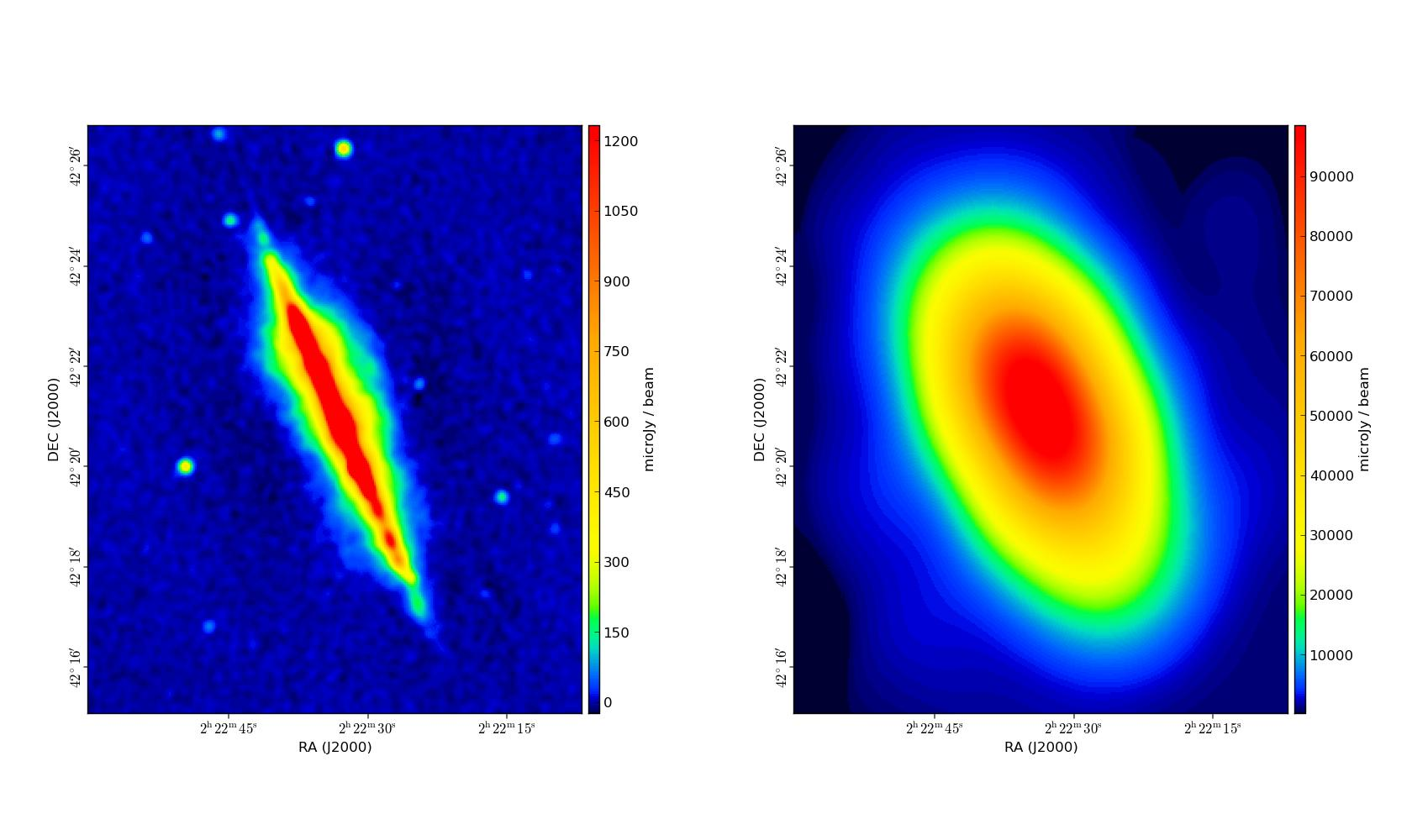}
 \caption{VLA map (left) of the edge-on galaxy NGC~891. The map shows a missing flux that can be filled in with a single-dish Effelsberg map (right)
 \cite{schmidt+17}}
 \label{imerg1}
\end{figure}

\begin{figure}[h]
 \includegraphics[scale=0.35,keepaspectratio=true,clip=true,trim=30pt 0pt 0pt 0pt]{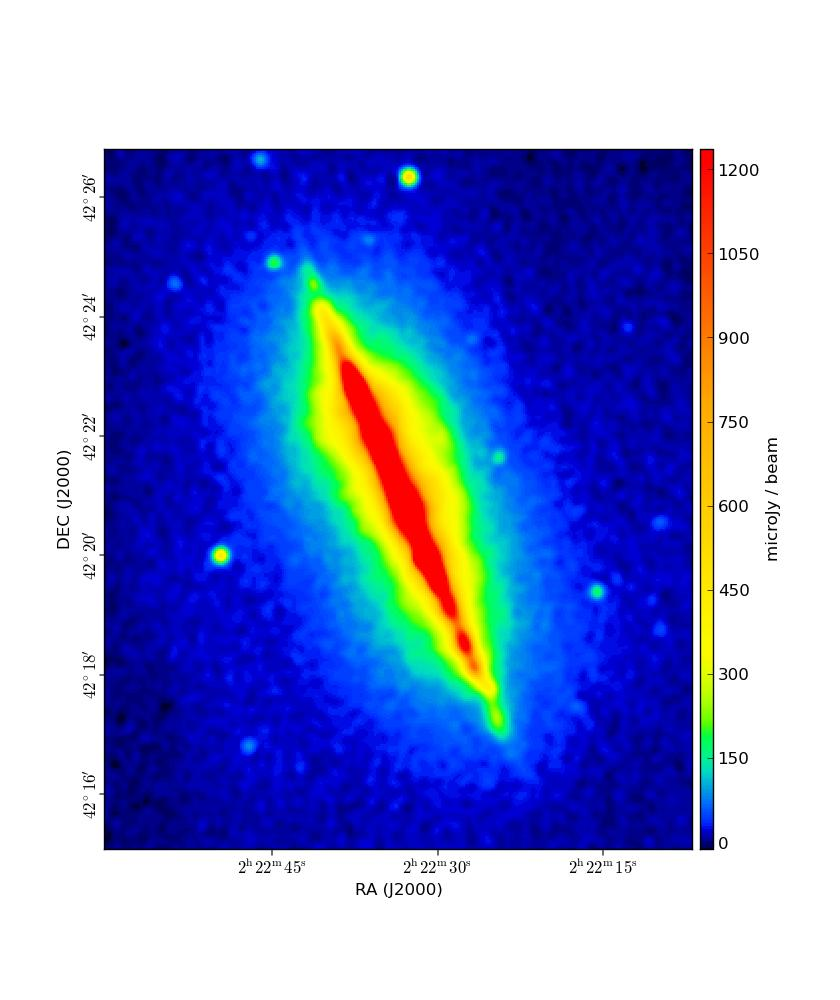}
 \caption{Combination of the VLA map and Effelsberg maps of NGC~891 \cite{schmidt+17}}
 \label{imerg2}
\end{figure}

The single-dish map has to be resampled to the pixel resolution of the interferometer map and the interferometer
map has to be convolved to the angular resolution of the single-dish map. The difference of both maps gives the missing flux
of the interferometer map with the angular resolution of the single-dish map. In order to add the difference map
to the interferometer map it is necessary to deconvolve the difference map to the resolution of the interferometer
map.

The spatial structure of the difference map contains no structures smaller than the resolution of the single-dish
map. Given that there is an overlap in the spacial frequency range between the single-dish and interferometer
maps, the error should be small if we approximate the deconvolution by simply rescaling the difference map
with the ratio of their beam areas $\Omega$,
\begin{equation}
 \begin{array}{rcl}
  D(l, b) &=& SD(l, b) - I(l, b) \, ,\\
  d(l, b) &=& [\Omega_{\rm I}/\Omega_{\rm SD}] \cdot D(l, b) \, ,\\
  \hat{I}(l, b) &=& I(l, b) + d(l, b) \, ,
 \end{array}
\end{equation}
where $D(l, b)$ is the difference map between the single-dish map $SD(l, b)$ and the interferometer map $I(l, b)$
with their beam areas $\Omega_{\rm I}$ and $\Omega_{\rm SD}$. The rescaled map $d(l, b)$ can now be added to the
interferometer map $I(l, b)$.

The advantage of this method is that the missing flux can be determined directly from the difference of the two maps
in the image domain.

\subsection{Box integration of edge-on galaxies: {\it BoxModels}}

The determination of the vertical scale heights of the total radio continuum
intensity of spiral galaxies seen edge-on (applicable for galaxy inclinations $>~80\degr$)
goes back to the method initially described and applied by Dumke et al, (1995).
The task is called 'BoxModels' and works with
(total intensity) maps in FITS format smoothed to a circular beam.

Within the task, strips are defined as perpendicular to the major axis of the galaxy (called the
z-direction) by giving its position angle. The width and lengths of each strip can
be defined with respect to the central position of the galaxy. Within one strip, the number and width
of boxes can be chosen wherein the radio intensities are averaged. As errors of these values the noise errors
within one box can be taken. The better way is to take the standard deviation of the mean in each box. In order to
exclude the expected intensity gradient in the z-direction, the standard deviation is calculated along each rows parallel
to the major axis of the galaxy separately within each box. Finally, they are averaged over all rows within each box.
These averaged intensity distributions along each strip are used to fit the vertical radio scale height (along each single strip).

For inclinations $i < 90\degr$, the observed intensity distribution is the superposition of the
intrinsic vertical distribution of the disk and halo emission and the inclined disk emission (even for an infinitesimally
thin disk), convolved with the telescope beam. Because we are only interested in the
vertical intensity distribution, the observed distribution has to be corrected for the contribution of the disk, which depends only on the diameter of the disk and
its inclination.

We approximately compensate the projection of the inclined galaxy with an increase of the beam width perpendicular to its disk by
$\Delta{HPBW} = R \cos(r/R \cdot \pi/2) \cos{i}$,
where $HPBW$ is the half power beam width, $R$ is the radius of the galaxy, $r$ the distance from the centre, and $i$ its inclination.
Then the effective $HPBW_{\rm eff} = \sqrt{HPBW^2 + \Delta{HPBW}^2}$.
The resulting distribution is still Gaussian in the vertical direction with
$\sigma_{\rm eff} = HPBW_{\rm eff}/(2\sqrt{2\ln{2}})$,
decreasing from the centre outwards along the major axis of the galaxy towards the telescope beam width $HPBW$ at the edges.

As an exact deconvolution of the observed intensity distribution with  $HPBW_{\rm eff}$ cannot be applied, we \emph{assume}
different possible vertical distributions convolved with the corresponding $HPBW_{\rm eff}$ for each strip and compare these
with the observed distributions by least-squares fits. Similar to Dumke et al (1995), an intrinsic exponential

\begin{equation}
  w_{\mathrm{exp}}(z)=w_{0}\exp{(-z/h)}\,,
\end{equation}
or an intrinsic  Gaussian distribution

\begin{equation}
  w_{\mathrm{gauss}}(z)=w_{0}\exp{(-z^{2}/h^{2})}\,
\end{equation}
is assumed with peak intensity $w_{0}$ and scale height $h$. This is convolved with the
effective telescope beam
$HPBW_{\rm eff} = 2\sqrt{2\ln{2}}\,\sigma_{\rm eff}$ ,

\begin{equation}
\label{beam}
  g(z)=\frac{1}{\sqrt{2\pi\sigma_{\rm eff}^{2}}}\exp{(-z^{2}/2\sigma_{\rm eff}^{2})}\,,
\end{equation}
which describes the contribution of the telescope beam and the projected
emission of the infinitesimally thin disk.
The convolved emission profile has the form

\begin{eqnarray}
\label{convexp}
  W_{\mathrm{exp}}(z)=\frac{w_{0}}{2}\exp{(-z^{2}/2\sigma_{\rm eff}^{2})} \cdot \nonumber\\
  \left[\exp{\left(\frac{\sigma_{\rm eff}^{2}-zh}{\sqrt{2}\sigma_{\rm eff} h}
  \right)}^{2}\mathrm{erfc}\left(\frac{\sigma_{\rm eff}^{2}-zh}
  {\sqrt{2}\sigma_{\rm eff} h}\right)\right.+ \nonumber\\
  \left.\exp{\left(\frac{\sigma_{\rm eff}^{2}+zh}{\sqrt{2}\sigma_{\rm eff} h}
  \right)}^{2}\mathrm{erfc}\left(\frac{\sigma_{\rm eff}^{2}+zh}
  {\sqrt{2}\sigma_{\rm eff} h}\right)\right]\,,
\end{eqnarray}

\begin{figure}[h]
 \includegraphics[scale=0.35,keepaspectratio=true,clip=true,trim=30pt 0pt 0pt 0pt]{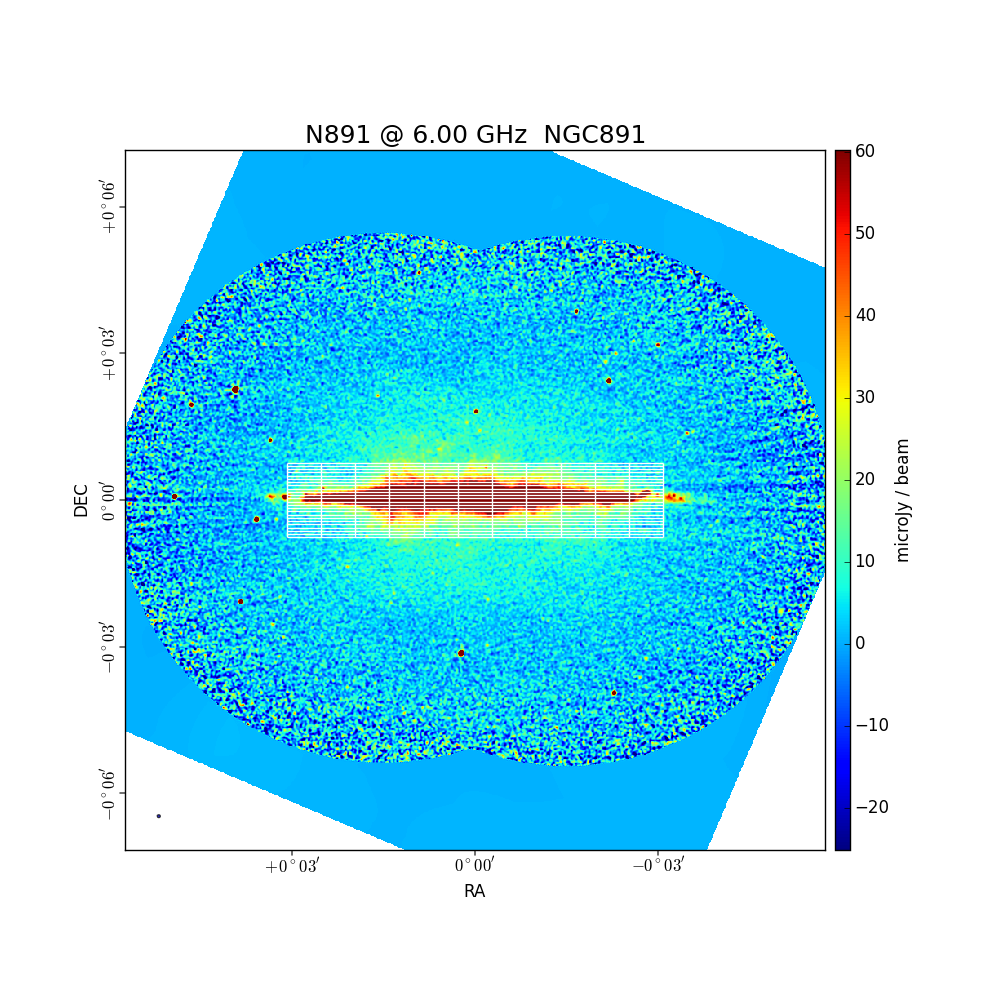}
 \caption{Combination of the VLA and Effelsberg maps, rotated by the position angle and boxes indicated.}
 \label{ngx891pix}
\end{figure}
\begin{figure}[h]
 \includegraphics[scale=0.35,keepaspectratio=true,clip=true,trim=30pt 0pt 0pt 0pt]{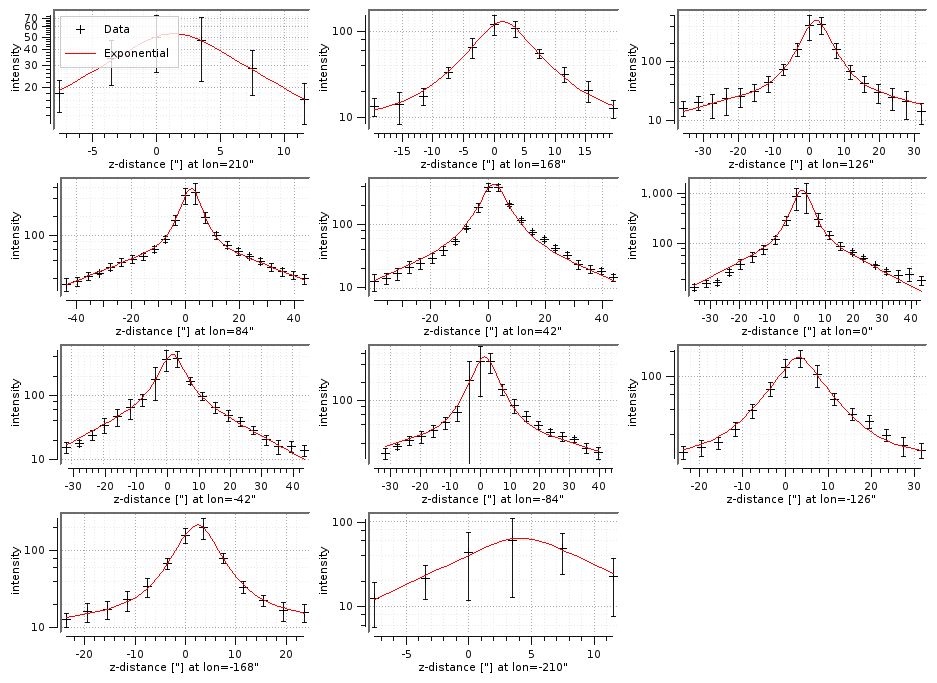}
 \caption{For each individual strip along the major axis a profile is created from which a two-component scale-height model is fitted.}
 \label{ngx891graph}
\end{figure}
in case of an exponential intrinsic distribution \cite{sievers88}, where
$\mathrm{erfc}$ is the complementary error function, defined as
\begin{equation}
 \mathrm{erfc}\,x=\frac{2}{\sqrt{\pi}}\int_{x}^{\infty}\!\exp{(-r^{2})}\,\mathrm{d}r\,.
\end{equation}

For a Gaussian intrinsic distribution, the convolution yields
\citep{dumke94}
\begin{equation}
\label{convgauss}
 W_{\mathrm{gauss}}(z)=\frac{w_{0}h}{\sqrt{2\sigma_{\rm eff}^{2}+h^{2}}}\exp{(-z^{2}/(2\sigma_{\rm eff}^{2}+h^{2}))}\,.
\end{equation}

The vertical radio scale heights $h$ can be determined by fitting one or two  'convolved' exponential or Gaussian functions of the
form given in Eq.~\ref{convexp} or Eq.~\ref{convgauss} to the averaged  radio intensities along each strip of a galaxy. The least-squares fits are made using the $Levenberg-Marquard$ method. Single boxes along a strip with mean values that are smaller than the rms-noise
intensity level of the map are omitted by the fitting procedure. The quality of the fit along each strip is determined by a reduced
$\chi2$-test.

\subsection{Integration of galaxy segments in face-on view}
In addition to the integration within boxes (as decribed above) and
elliptical rings, NOD3 also offers the possibility to integrate map
values in equally sized segments of rings that are circular in the plane of the galaxy. The primary
application of this is to study different regions of spiral galaxies in
a face-on view even if they are inclined. For this purpose, the
task first rotates the input map by a user-given position angle and
then deprojects the map by the specified inclination angle of the
galaxy, so that it appears face-on. This is possible for inclinations up
to $80\degr$.

\begin{figure}[h]
 \includegraphics[scale=0.35,keepaspectratio=true,clip=true,trim=30pt 0pt 0pt 0pt]{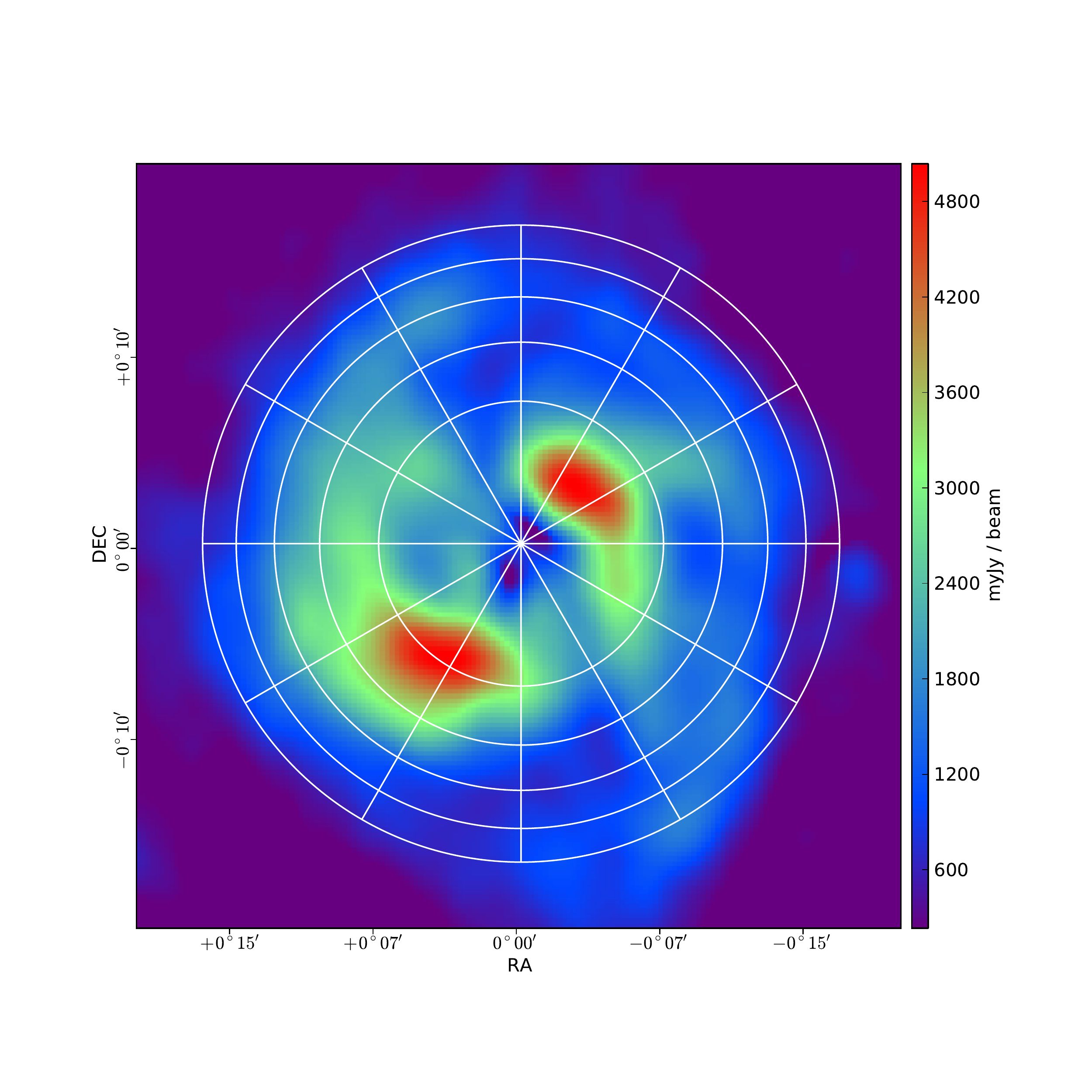}
 \caption{Example of a sectored target.}
 \label{fig:galsec}
\end{figure}
\begin{figure}[h]
 \includegraphics[scale=0.35,keepaspectratio=true,clip=true,trim=30pt 0pt 0pt 0pt]{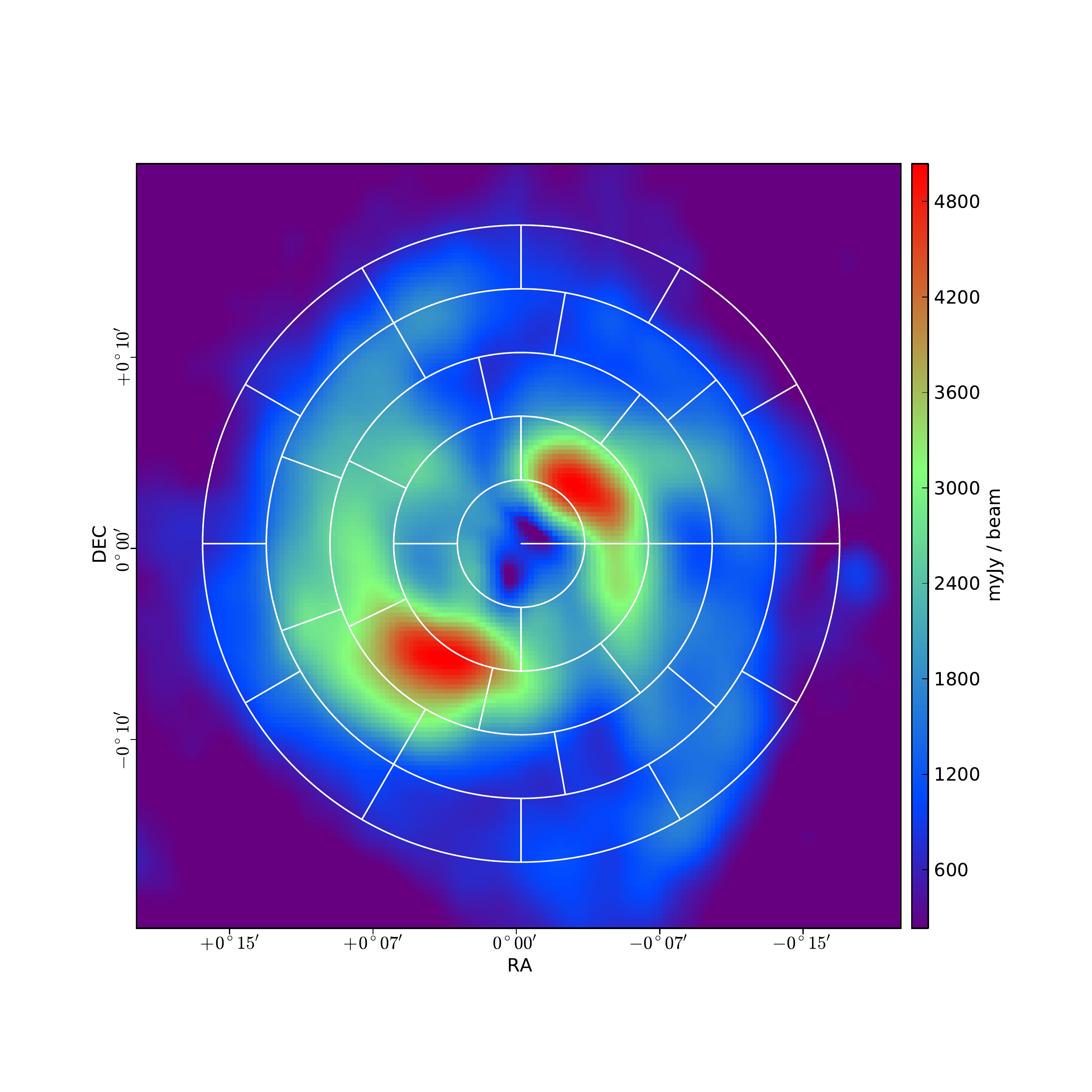}
 \caption{Example of a segmented target.}
 \label{fig:galseg}
\end{figure}

The integrated and averaged values are then computed for each
segment in a set of concentric rings, where the number of rings (up to
15), the number of segments (up to 180 per ring), and the maximum ring
radius are selected by the user. For the arrangement of the ring
segments, two different options are offered. (i) In sector mode, all rings are divided
into an equal number of segments with a constant separation in azimuth,
in which case the width of the rings decreases with radius
(Fig.~\ref{fig:galsec}). (ii) In segment mode, all rings have the same
width, so that the number of segments per ring increases with radius
(Fig.~\ref{fig:galseg}).


\section{Summary}

The NOD3 program package is designed to reduce and analyse single-dish maps. It is the result of developing an interactive software package with a
GUI that includes all the necessary programs to obtain the final maps
in total intensity and linear polarisation. The program package is written in Python. The required data format for the 
input maps is FITS.
The NOD3 package is especially powerful to remove 'scanning effects'
due to clouds, receiver instabilities, or RFI in single-dish observations.
We significantly improved or revised several of the data reduction methods introduced by NOD2.
The 'basket-weaving' tool combines orthogonally scanned maps to a final map that is almost free of scanning effects.
The new restoration tool for dual-beam observations reduces the noise by a factor of about two.
The combination of single-dish with interferometer data in the map plane ensures the recovery of the full flux density.
The NOD3 package also offers an improved method for bias correction of the polarised intensity map {}\cite{mueller+17}, and it offers the possibility of including special tasks written by the individual observer.

The NOD3 package is designed to be extendable to multi-channel data represented by data cubes in Stokes I, Q, and U,
which is planned as the next step of our work. Another future extension will be the cleaning of the telescope side lobes.

The NOD3 software package is available under the open source license GPL for a free use at other single-dish
radio telescopes of the astronomical community.
The latest version can be obtained at the $gitlab$ server of MPIfR Bonn at \\

https://gitlab.mpifr-bonn.mpg.de/peter/nod3-single-dish-reduction-software \\

A NOD3 user manual is included and available from the $gitlab$ at the MPIfR Bonn.

\begin{acknowledgements}
      We thank Ren\'e Gie{\ss}\"ubel, Maja Kierdorf, and Carolina Mora for testing
      the program package, Ren\'e Gie{\ss}\"ubel for his work on the
      NOD3 User Manual, and Uli Klein for fruitful discussions. We
      also thank Alexander Kraus for critical comments on the manuscript.
\end{acknowledgements}

\appendix

\section{Toolbox: The pre-processing software at the Bonn 100-m Telescope}

The data observed in the mapping mode (``on the fly'') with the Effelsberg 100-m telescope are stored as they are received from the backends together with
telescope position information and their time stamps. This is necessary to link astronomical coordinates and flux intensities. The mapping mode allows coverage of an
arbitrarily chosen area on the sky that is measured in astronomical coordinates. The coverage can be scanned along the longitude  axis or latitude axis in any 
coordinate system. The non-scanning
coordinate is always fixed for one scan. With the following scans the non-scanning coordinate is incremented usually by a third of the telescope beam width 
(sampling theorem).

All frontends and backends at the 100-m radio telescope at Effelsberg are steered by uniform $blank$ and $sync$ signals produced by hardware.
A blank-sync generator is
designed to produce short time signals interrupted by a blank time signal. After $n$ blank time signals a sync time signal is sent that contains the information of the
integration time between the blank time signals. The length of a sync time signal is called $phase time$, i.e. the integration time of one $phase$. The blank time is dead
time while the measurement is not recording. The blank time is much smaller then the sync time. Within one $phase$ a noise diode or any other signal could switch
alternately on and off to the frontend. We call a successive series of same phases $phase\, cycle$. A $phase\, cycle$ is used for a pre-reduction in order to calibrate
the astronomical signals.

Each $phase$ is written to disk in the $MBFITS$ format (Muders et al. 2017) during the observations. This data is used by a dedicated preprocessing software
called $Toolbox$, which, similar to NOD3, is completely written in Python. The $MBFITS$ files are used as input for the Toolbox. All radio continuum and spectral
polarimeter data are pre-reduced by the Toolbox. Preprocessing means mainly calibrating the astronomical signals by the signal of noise diode ($cal\, signal$)
removing radio frequency interference $RFI$, pre-adjusting the baselines, preparing, and calibrating the polarisation data.

The Toolbox is also used for the analysis of, for example cross-scans over a point-like source,
on-offs, or skydips. In the case of sky mapping, the Toolbox outputs a FITS file ready for the NOD3 software.

\end{document}